# Kilowatt-level Yb-Raman fiber amplifier with narrow-linewidth and near-diffraction-limited beam quality

**Pengfei Ma, Yu Miao, Wei Liu, Daren Meng, and Pu Zhou**

College of Advanced Interdisciplinary Studies, National University of Defense Technology, Changsha, Hunan, 410073, People's Republic of China

E-mail: aiken09@163.com

**Abstract:**
By focusing on a typical emitting wavelength of 1120 nm as an example, we present the first demonstration of a high-efficiency, narrow-linewidth kilowatt-level all-fiber amplifier based on hybrid ytterbium-Raman (Yb-Raman) gains. Notably, two temporally stable, phase-modulated single-frequency lasers operating at 1064 nm and 1120nm, respectively, were applied in the fiber amplifier, to alleviate the spectral broadening of the 1120 signal laser and suppress the stimulated Brillouin scattering (SBS) effect simultaneously. Over 1 kW narrow-linewidth 1120 nm signal laser was obtained with a slope efficiency of ~ 77% and a beam quality of $M^2$~1.21. The amplified spontaneous emission (ASE) noise in the fiber amplifier was effectively suppressed by incorporating an ASE-filtering system between the seed laser and the main amplifier. Further examination of the influence of power ratios between the two seed lasers on the conversion efficiency had proved that the presented amplifier could work efficiently when the power ratio of 1120 nm seed laser ranged from 31% to 61%. Overall, this setup could provide a well reference for obtaining high power narrow-linewidth fiber lasers operating within 1100-1200 nm.

## 1. Introduction

High power narrow-linewidth fiber lasers operating at 1100-1200 nm have been highly desired for second harmonic generation to further obtain laser sources operating at the yellow-green spectral region [1-6]. Beneficial from the wide gain spectra of ytterbium (Yb) ions in the silica host (976-1200 nm) [7], directly using sole Yb-ions gain is a simple approach to achieve narrow-linewidth emissions at longer wavelength [2, 8-10]. However, due to the strong amplified spontaneous emission (ASE) and the following destructive self-oscillation [11], this approach has becoming sparse with further power scaling. To overcome the limitations mentioned above, active fiber with different dopant mechanism and special structure designs were performed previously [12-15]. Except for power scaling from directly active fiber, as one of the prevailing techniques, narrow-linewidth Raman fiber amplifiers (RFAs) have also been proposed and widely applied in the previous works [1, 3, 4, 16, 17]. Notably power handling ability of wavelength division multiplexer (WDM) and the stimulated Brillouin scattering (SBS) effect are the two majority limitations for their power scaling [16-20]. Despite several techniques had been provided, such as acoustically tailored fiber [16], longitudinally varied strain [18, 19], cascaded Raman gain fiber [20], the output power of narrow-linewidth RFAs emitting within wavelength range of 1100-1200 nm has been still below hundred-watt [17, 19]. Notably, counter-pumped manner was previously thought as the necessary condition for suppressing spectral linewidth broadening in narrow-linewidth RFAs [1, 16-20]. Consequently, despite that hybrid Yb-Raman nonlinear fiber amplifier was proposed to break through the limitation of WDM in traditional

RFAs [21], co-pumped manner is necessary for amplification and conversion of Raman-pumped laser to the Raman signal laser, which seems to be unsuitable for narrow-linewidth Raman amplification because that obvious spectral broadening phenomenon was observed in the Raman signal laser during power amplification previously [21-23].

Quite recently, our theoretical analysis has been shown that the spectral broadening phenomenon in narrow-linewidth hybrid rare-earth-Raman fiber amplifiers is induced by the strong intensity noise in the Raman signal laser. This type of intensity noise could origin both from the intensity noises of inserted Raman seed laser and the Raman-pumped laser [24], which is quite different from the noise transfer characteristics of the direct Yb-doped fiber amplifiers (YDFAs). Accordingly, narrow-linewidth operation could even be maintained in co-pumped manner when both the inserted Raman seed laser and the Raman-pumped laser are temporally stable laser sources. This analysis provides a promising design concept for fulfilling high power, narrow-linewidth fiber lasers emitting within 1100-1200 nm by hybrid Yb-Raman gains, in which the limitation of WDM in previous narrow-linewidth RFAs could be avoided and conventional large-mode-area Yb-doped fiber with shorter effective length could be employed for favorable of SBS suppression. In the previous work, we firstly conducted a principle-of-concept experiment based on co-pumped pure single frequency RFA, and it is shown that the temporally stable Raman-pumped source could be fulfilled by cascaded power scaling of a low-noise, phase-modulated, single-frequency seed laser [25], which provides a robust way to generate multi-kilowatt Raman-pumped laser in practice [26, 27].

In this work, we report a high-efficiency, narrow-linewidth kilowatt 1120 nm all-fiber amplifier based on hybrid Yb-Raman gains. Two phase-modulated single-frequency lasers operating at 1120 nm and 1064 nm, respectively, were cascaded amplified to generate temporally stable inserted Raman signal laser and Raman-pumped laser for alleviating the spectral broadening phenomenon in the Raman signal laser and performing SBS suppressing for both the Raman-pumped laser and Raman signal laser. An ASE-filtering system was applied afore the main amplifier to weaken the ASE noise. At a 976 nm pump power of about 1.35 kW, narrow-linewidth kilowatt 1120 nm fiber laser was obtained with high slope efficiency operation (77%) and excellent beam quality ($M^2$~1.21). Further examinations real that the presented amplifier could still work efficiently when the power ratio of the 1120 nm inserted laser ranged from 31% to 61%.

## 2. Experimental results

Fig. 1 illustrates the experimental setup of the all-fiber amplifier system, which mainly consists of three parts: the two seeds, the pre amplifier with an ASE-filtering system, and the main amplifier. As shown in Fig. 1(a), each seed laser consisted of its own phase modulation and pre-amplification subsystems. In Seed1 (Raman-pumped seed), a low noise single-frequency laser with central wavelength of 1064 nm and output power of about 40 mW [28], was externally modulated through a phase modulator (PM1) driven by a white noise signal (WNS) to broaden its spectral linewidth for SBS suppression [26]. Then, the 1064 nm seed laser was power amplified by two conventional single mode YDFAs. In Seed2 (Raman signal seed), a single-frequency laser with central wavelength of 1120 nm and output power of about 27 mW was firstly externally modulated through a phase modulator driven by WNS to broaden the spectral linewidth and then power amplified by two conventionally single mode, counter-pumped Raman fiber amplifiers (RFAs). Two tap couplers with 0.33% and 1% backward power ratios were respectively spliced onto the output of Seed1 and Seed 2 to measure the backward propagating lasers of 1064 nm and 1120 nm and ensuring the safety of the two seeds.

As shown in Fig. 1(b), the output lasers from Seed1 and Seed2 were firstly combined through a wavelength division multiplexer (WDM1), and then coupled into the pre-amplifier through a pump-signal combiner (Combiner) together with two pump laser diodes (LDs) with the central wavelength of 976 nm. The active fiber was commercial double clad Yb-doped fiber (YDF), which had a core diameter of 10 μm and an inner cladding diameter of 125 μm. The cladding absorption coefficient of the active fiber was about 4 dB/m at 976 nm, and 20 m long active fiber was used here. Through adjusting the inserted seed powers (Seed1 and Seed2) and 976 nm pump power, we could flexibly adjust the output powers of the lasers operating at 1064 nm and 1120 nm in this pre-amplifier. At the maximum 976 nm pump power of about 172 W, the output powers of the lasers operating at 1064 nm and 1120 nm were about 23 W and 106 W, respectively. Thus, most of the pump laser could be converted into the 1120 nm signal laser in this pre-amplifier based on hybrid Yb-Raman gains. In addition, no spectral broadening phenomenon was observed for the 1120 nm signal laser during amplification in this strictly single-mode pre-amplifier.

To weaken the ASE noise, the output laser of the pre-amplifier was coupled into an ASE-filtering system, which mainly consisted of two WDMs, two band-pass filters (BPFs), and two couplers. In the ASE-filtering system, the lasers operating at 1064 nm and 1120 nm were first separated by the WDM2. Then the two lasers passed through the BPFs with the central wavelength of 1064 nm (BPF1) and 1120 nm (BPF2), respectively, to filter the ASE noise. Finally, the two lasers were combined through the WDM3. The bandwidths of the two BPFs were both about ± 5 nm. Two 1% tap couplers were applied here to measure the backward propagating lasers from the main amplifier. Here, we set the output powers of Seed1 and Seed2 both to be about 1.2 W, and the 976 nm pump power of the pre amplifier to be about 55 W. Then, the output powers of the lasers operating at 1064 nm and 1120 nm after the ASE-filtering system were about 11 W (39%) and 17 W (61%), respectively.

As shown in Fig. 1(c), the output laser after the ASE-filtering system (Seed3) was coupled into the main amplifier through a mode field adaptor (MFA) and a pump combiner. As for the main amplifier, six high power LDs (central wavelength~976 nm) were incorporated into a (6+1)×1 pump combiner to pump the active fiber. The active fiber was commercial double clad YDF, which had a core diameter of 20 μm and an inner cladding diameter of 400 μm. The cladding absorption coefficient of the active fiber was about 1.2 dB/m at 976 nm, and 18 m long active fiber was used in the amplifier. The amplifier output end was terminated with a cladding mode stripper (CLS) and a quartz beam hat (QBH). To separate the 1120 nm signal laser from the residual laser operating at 1064 nm, a dichroic mirror (DM) was applied, then the properties of the output lasers at 1064 nm and 1120 nm could be measured at Port A and Port B, respectively.

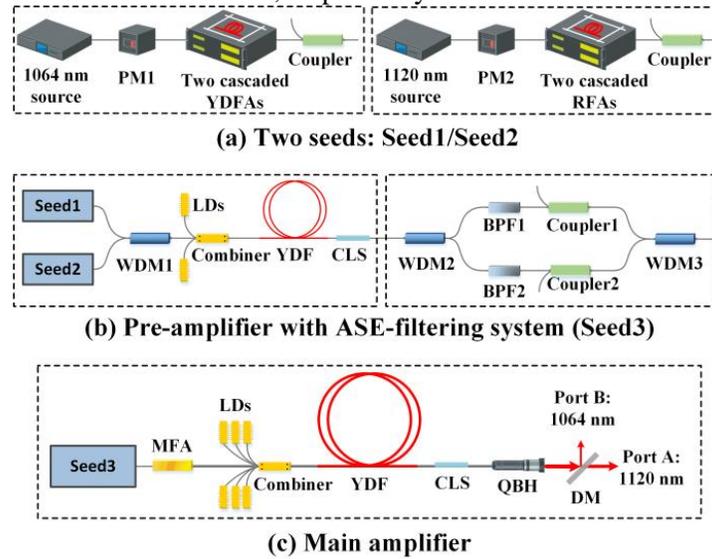

Fig. 1. Experimental setup for the kilowatt 1120 nm fiber amplifier

Before demonstrating the main experimental results, we would show the contributions of the ASE-filtering system through a contrast experiment by removing the ASE-filtering system in the experimental setup. Fig. 2 illustrates the total normalized output spectra of the amplifier at a 976 nm pump power of about 1 kW, when the ASE-filtering system was removed or applied. As for the laser at 1064 nm, the suppression levels of the ASE noise were about 19 dB and 38 dB for the two cases, respectively. As for the laser at 1120 nm, the suppression levels of the ASE noise were about 30 dB and 49 dB for the two cases, respectively. Thus, the ASE-filtering system led to a decrease of the ASE noise by about 19 dB for the amplifier at a 976 nm pump power of about 1 kW. Accordingly, the ASE-filtering system could effectively suppress the ASE noise and improve stability of the fiber amplifier.

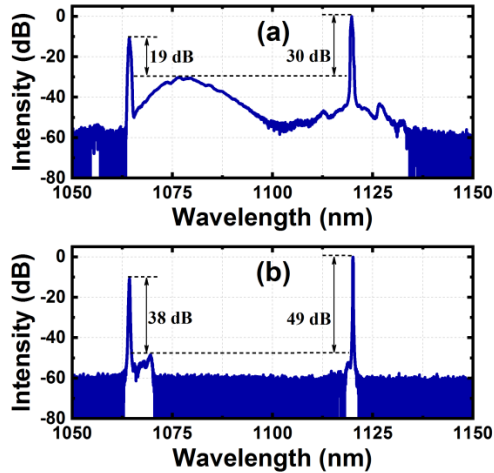

**Fig. 2.** Total normalized output spectra of the amplifier: (a) the ASE-filtering system was removed; (b) the ASE-filtering system was applied

Fig. 3 illustrates the output powers of different spectral components in the main amplifier. As shown in Fig. 3, both the output powers of lasers at 1064 nm and 1120 nm grew with the 976 nm pump power below 0.54 kW. When the pump power exceeded 0.6 kW, the output power of 1120 nm signal laser kept growing while output power of laser at 1064 nm began to decrease. At the maximum power pump of about 1.35 kW, the corresponding output powers of lasers at 1120 nm and 1064 nm were about 1.04 kW and 0.05 kW, respectively, thus the power ratio of 1120 nm signal laser was as high as about 95.7%. The power transferring efficiency from the 976 nm pump laser to the 1120 nm signal laser at 1.04 kW was calculated to be 77%. It should be mentioned that there was no sign of nonlinear increase for the power of backward propagating laser at 1120 nm or 1064 nm, which means that the fiber amplifier worked below the SBS threshold at a pump power of about 1.35 kW.

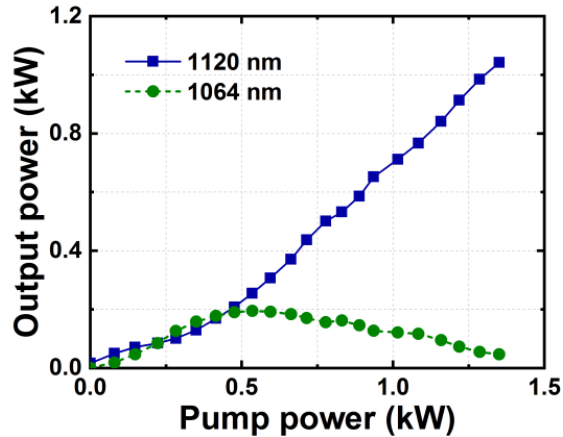

**Fig. 3.** Measured output powers at different 976 nm pump powers

The integral output spectrum and the output spectrum for the signal laser at 1120 nm at the maximum output power were measured through a commercial optical spectral analyzer (OSA), which is shown in Fig. 4. As shown in Fig. 4(a), the suppression levels of the ASE noise for the laser at 1064 nm and 1120 nm are about 38 dB and 51 dB, respectively. It also revealed that the ASE noise was effectively suppressed in our experiment. As shown in Fig. 4(b), the output laser at Port A was pure 1120 nm signal laser, thus the measured output powers for the 1120 nm signal laser in Fig. 2 were reliable. The inset figure in Fig. 4(b) illustrates the laser beam profile at 1 kW signal power. The beam quality result indicates that near diffraction-limited operation is achieved with the M2 value of about 1.21.

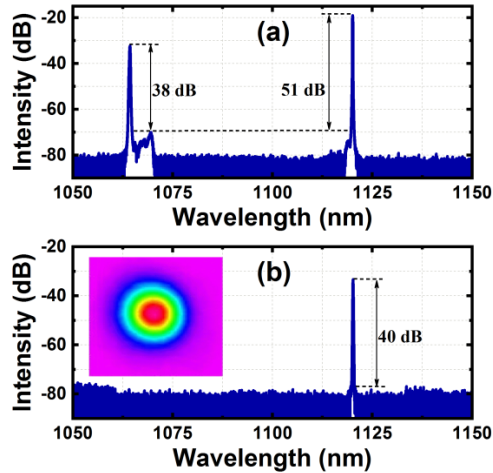

**Fig. 4.** Measured output spectra at the maximum output power: (a) the integral output spectrum; (b) the output spectrum at Port A

To ensure enough enhancement of the SBS threshold for the fiber amplifier, the full width at half maximum (FWHM) spectral linewidth of the 1120 nm seed laser was broadened to about 68 pm before injecting into the main amplifier. Fig. 5 illustrates the FWHM spectral linewidths of the 1120 nm signal laser. As shown in Fig. 5, the FWHM spectral linewidth of the 1120 nm signal laser nearly kept around 68 pm when the output power was below 0.71 kW. However, when the output power reached 0.84 kW, the FWHM spectral linewidth of the 1120 nm signal laser got broadened to about 96 pm abruptly. The FWHM spectral linewidth at the maximum output power was about 97 pm, and the corresponding spectral broadening factor was just about 1.4 compared to the inserted 1120 nm seed laser. In fact, this phenomenon mainly corresponds to the spectral broadening of the wing part, because the 10 dB spectral linewidth of 1120 nm signal laser gradually got broadened from the initial 136 pm to about 180 pm when the output power increased to 0.71 kW. This spectral broadening phenomenon in the hybrid Yb-Raman is interesting, and deserves further investigation. The possible mechanisms were that the spectral filtering process or the ASE noise in the main amplifier might change the temporal property of fiber lasers, which could lead to the spectral broadening of the 1120 nm signal laser here [29]. More detailed explanations would require the overall consideration of the complex interactions among the ASE noise, the nonlinear effects and the gain dynamics [24, 30].

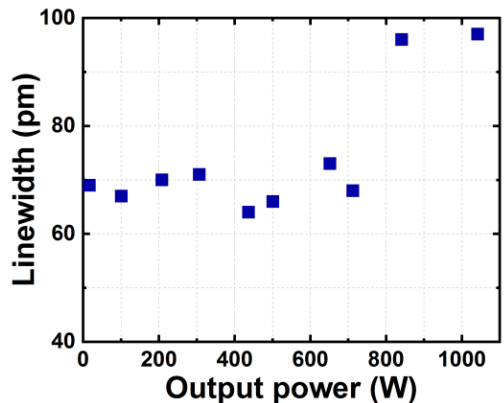

**Fig. 5.** Measured FWHM linewidths of the 1120 nm signal laser

To investigate the power transferring process between the lasers at 1064 nm and 1120 nm, it is necessary to investigate the possible influence of the power ratios of the two seed lasers on the fiber amplifier. Thus, we conducted control experiments through changing the power ratio of the 1120 nm seed laser when the total power of two seed lasers was kept nearly unchanged. Fig. 6 compares the output powers and the power ratios of the 1120 nm signal laser at different 976 nm pump powers when the injected power ratio of the 1120 nm was about 31%,

43%, 52%, and 61%, respectively. As shown in Fig. 6(a), the variation trends of the output power were similar to each other. And the maximum output powers of the 1120 nm laser were measured to be about 1009 W, 1037 W, 1044 W, and 1042 W, respectively. As shown in Fig. 6(b), the variation trends for the power ratio of the 1120 nm signal laser were also similar to each other. And the ultimate 1120 nm power ratios were calculated to be 93.1%, 94.9%, 95.3% and 95.7%, respectively, which was gradually optimized a little in the experiment. Compared with other three cases, the ultimate 1120 nm power ratio just fell by 1.8%~2.6% when the injected power ratio of the 1120 nm seed laser was about 31%. Further reducing the power ratio of the 1120 nm seed laser would lead to the more decrease of the output power and power ratio. Besides, further increase of the power ratio of 1120 nm seed laser, the abnormal increase of backward power is observed by detecting the backward port of coupler inserted in Seed 1, which denotes that the ASE noise made the whole amplification system fragile in this case. Overall, our amplifier could work with high efficiency beyond kilowatt-level output power when the power ratio of the 1120 nm seed laser ranged from 31% to 61%.

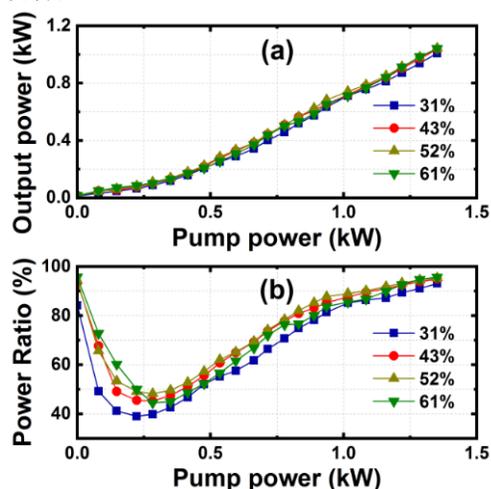

**Fig. 6.** Properties of the signal: (a) output power; (b) power ratio

### 3. Conclusions

In conclusion, power scaling of a narrow-linewidth all-fiber amplifier operating at 1120 nm based on hybrid Yb-Raman is presented. Two temporally stable, phase-modulated single-frequency lasers operating at 1064 nm and 1120 nm, respectively, were employed as the hybrid seed to alleviate the spectral broadening and balance SBS effect. An ASE-filtering system was applied to weaken the ASE noise and improving stability of the fiber amplifier. Narrow-linewidth (97 pm) kilowatt 1120 nm signal laser was obtained with high slope efficiency (77%) and excellent beam quality ($M^2$~1.21). This fiber amplifier could work efficiently when the power ratio of the 1120 nm seed laser ranged from 31% to 61%. Significantly, the overall presentation could provide instructive suggestions for the generation of high brightness, narrow-linewidth fiber lasers emitting within the 1100-1200nm long wavelength range.